\documentclass[aps,prc,twocolumn,showkeys,showpacs]{revtex4-1}
\usepackage{graphicx}

\begin{document}
\title
   {\bf Spontaneous fission half-lives of nuclei in a phenomenological model}
\author{A. Zdeb, M. Warda, and K. Pomorski}
\email{Krzysztof.Pomorski@umcs.pl}
\affiliation{Katedra Fizyki Teoretycznej, Uniwersytet Marii
             Curie-Sk\l odowskiej, Lublin, Poland}
\date{\today}

\begin{abstract}
A simple phenomenological model, based on the \'Swi{\c a}tecki idea for
evaluation of the spontaneous fission half-lives is proposed. The model contains
only one adjustable parameter fixed to the data for even-even nuclei and two
additional hindrance factors to the life-times, which give the effect of odd
particles. A good agreement with the experimental data for all fissioning nuclei
is achieved.
\end{abstract}
\maketitle

\section{Introduction}

\begin{figure}[h!]
 \includegraphics[height=0.8\columnwidth,angle=270]{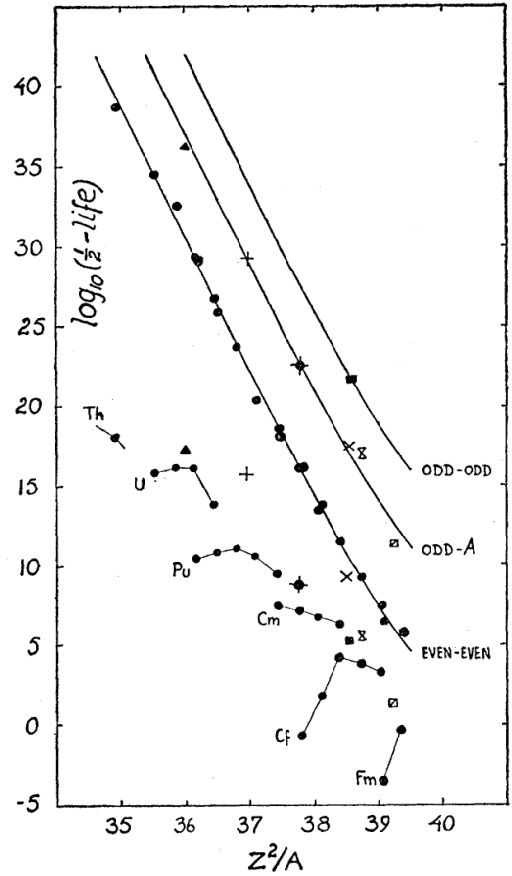}
\caption{Experimental (bottom part) and corrected by the ground-state shell
effects (EVEN-EVEN, ODD-A and ODD-ODD) spontaneous fission half-lives of atomic
nuclei as a function of fissility parameter (original figure from Ref.
\cite{Sw55}.}
\label{fig1}
\end{figure}

Close correlations between spontaneous fission half-lives and ground
state masses of nuclei were noticed in 1955 by W. J. \'Swi{\c a}tecki
\cite{Sw55}. He has proposed a simple formula, which has joined the
observed fission lifetimes with the difference between experimental and
liquid drop masses ($\delta M$). A regular dependence of
$\log_{10}T^{\rm sf}_{1/2}$ obtained by him after adding an empirical
correction proportional to $\delta M$ is presented in Fig.~\ref{fig1}
taken from Ref.~\cite{Sw55} as function of the fissility parameter
$Z^2/A$.

The aim of the present paper is to check if this 58 years old \'Swi{\c a}tecki
brilliant idea still works. A modern version of the liquid drop model derived
in Ref.~\cite{PD03} and all up to now measured spontaneous fission half-lives
\cite{nudat2} are taken in our analysis analysis.

\section{The liquid drop model}

The experimental ground-state masses of all nuclei with the proton number $Z\ge
8$ and neutron number $N\ge 8$ were calculated in Ref.~\cite{PD03} using the
macroscopic-microscopic model. The macroscopic part of this Lublin-Strasbourg
Drop (LSD) mass formula (in MeV units) was following:
\begin{equation}
\begin{array}{l}
M_{\rm LSD}(Z, N, {\rm def})=\\[+0.5ex]
~~ 7.289034\cdot Z+8.071431\cdot N-0.00001433\cdot Z^{2.39}\\[+0.5ex]
~ -15.4920(1 - 1.8601 I^2) A\\[+0.5ex]
~ + 16.9707(1 - 2.2938 I^2) A^{2/3}\, B_{\rm surf}({\rm def})\\[+0.5ex]
~ + 3.8602(1 + 2.3764 I^2) A^{1/3}\,B_{\rm cur}({\rm def})\\[+0.5ex]
~ + 0.70978\,Z^2/A^{1/3}\,B_{\rm Coul}({\rm def})
   - 0.9181\,Z^2/A\\[+0.5ex]
~ -10\exp({-4.2|I|})\,B_{\rm cong}({\rm def}) \,\,.
\end{array}
\label{LSD}
\end{equation}
while the ground state microscopic corrections $E_{\rm micro}$ were taken from
the mass tables of M\"oller at al. \cite{MN95}. In Eq.~(\ref{LSD}) $A=Z+N$
denotes the mass number, $I=(N-Z)/A$ reduced isospin and $B_{\rm surf}$, $B_{\rm
cur}$, $B_{\rm Coul}$ and $B_{\rm cong}$ are relative to the sphere: surface,
curvature, Coulomb and congruence (see Ref.~\cite{MS97}) energies. The
parameters in the first and the last row in Eq.~(\ref{LSD})are taken from
Ref.~\cite{MN95}, while the rest 8 parameter were fitted to the data.

It was shown in Refs. \cite{PD03,DP07,IP09} that the LSD model (\ref{LSD}),
which parameters were fitted to the experimental ground-state masses only, is
able to reproduce well the fission barrier heights of light, medium and heavy
nuclei when the microscopic part of the ground-state binding energy is taken
into account according to the topographical theorem of \'Swi{\c a}tecki
\cite{MS96}.

The liquid drop barrier height of actinides decreases almost linearly in
function of $Z$ from 4.3 MeV for $Z=90$ to 0 for $Z\ge 103$. The fission barrier
of finite heights appears in the super-heavy nuclei mostly due to the shell
effects in the ground state.

\section{Calculation details}

Following Ref.~\cite{Sw55} we have subtracted from the logarithm of the
spontaneous fission half-lives in years:
\begin{equation}
f(Z,N)=\log_{10} [T_{1/2}^{\rm sf}(Z,N)/y]+k\delta M(Z,N)\,\,.
\label{fZN}
\end{equation}
the {\it experimental} value of ground-state microscopic energy $\delta M(Z,N)$
can be evaluated using the experimental and the liquid drop (\ref{LSD}) mass
difference:
\begin{equation}
\delta M_{micr}^{\rm exp}(Z,N)=M_{\rm exp}(Z,N)-M_{\rm LSD}(Z,N,0)
\label{dM}
\end{equation}
multiplied by an arbitrary factor $k$.

\begin{figure}
 \includegraphics[height=\columnwidth,angle=270]{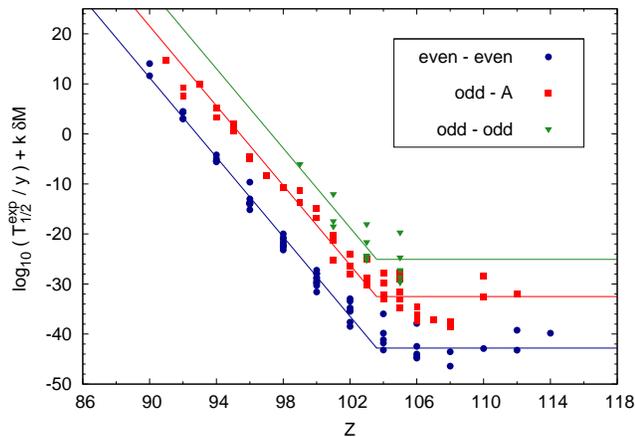}
\caption{ Logarithms of the observed spontaneous fission half-lives corrected
with masses "shifts" as a function of proton number.}
\label{fig2}
\end{figure}

In our analysis of function $f(Z,N)$ were evaluated for 62 even-even fissioning
isotopes with $Z\le 114$. Smooth dependence of $f(Z,N)$ on $Z$, in which shell
effects were almost wash out, was achieved for
$k=7.5$ MeV$^{-1}$ as one can see in Fig~\ref{fig2}.
For given element the dependence of the function $f(Z,N)$ on neutron number is
very week, so we omit this dependence in further consideration. The function
$f(Z)$ which represents the global trend in the fission life-times can be
approximated by two crossing straight lines:
\begin{equation}
f(Z)=\left\{
\begin{array}{ll}
 -4\cdot Z +371.2 ~~~&{\rm for~~} Z < 104\,\,,\\
 -42.8               &{\rm for~~} Z\ge 104\,\,.
\end{array}\right.
\label{fZ}
\end{equation}
The coefficients of these straight lines are simply given by the systematics
of points representing the corrected (according to Eq. (\ref{fZN})) fission
life-times. They are not free adjustable parameters.
The above dependence is very similar to the dependence of the liquid-drop
barrier heights in function of $Z$.
It is also visible in Fig.~\ref{fig2} that the slope of similar data for odd-$A$
and odd-odd nuclei is almost the same but the corresponding curves are shifted
by a constant. This shift, which can be called global hindrance factor is equal
to $h=10.3$ for odd-$A$ nuclei and 17.8 for odd-odd systems. It represents the
fission life-time increase given both by the odd-even effects in the ground
state energy and the fission barrier enhancement due to the angular momentum and
parity conservation of unpaired nucleons. Contrary to \'Swi{\c a}tecki
analysis \cite{Sw55} we assume here the linear in $Z$ form of the $f(Z)$ and we
have extrapolated it by a constant for the super-heavy nuclei.

\begin{figure}[bh!]
 \includegraphics[height=\columnwidth,angle=270]{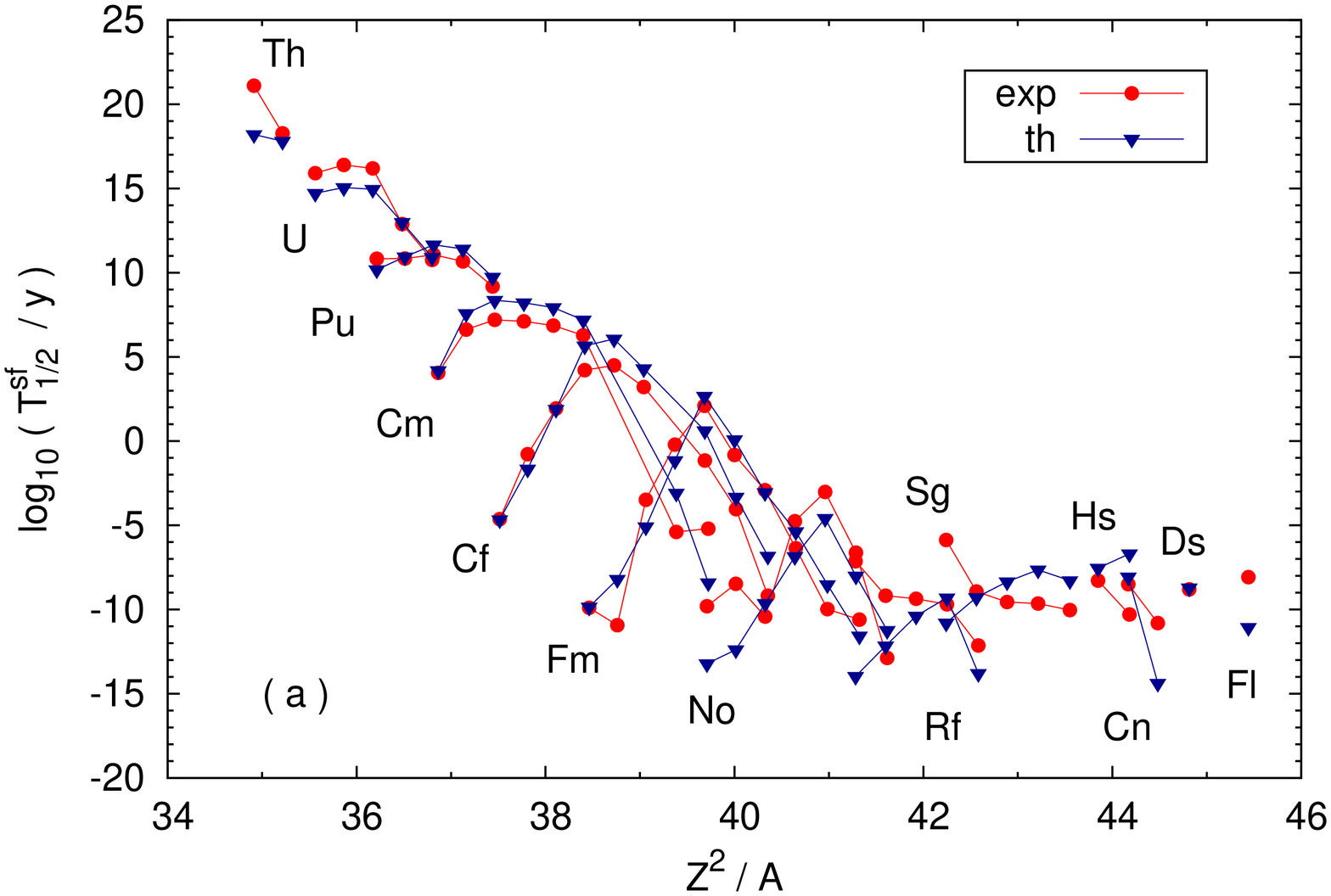}\\[-4ex]
 \includegraphics[height=\columnwidth,angle=270]{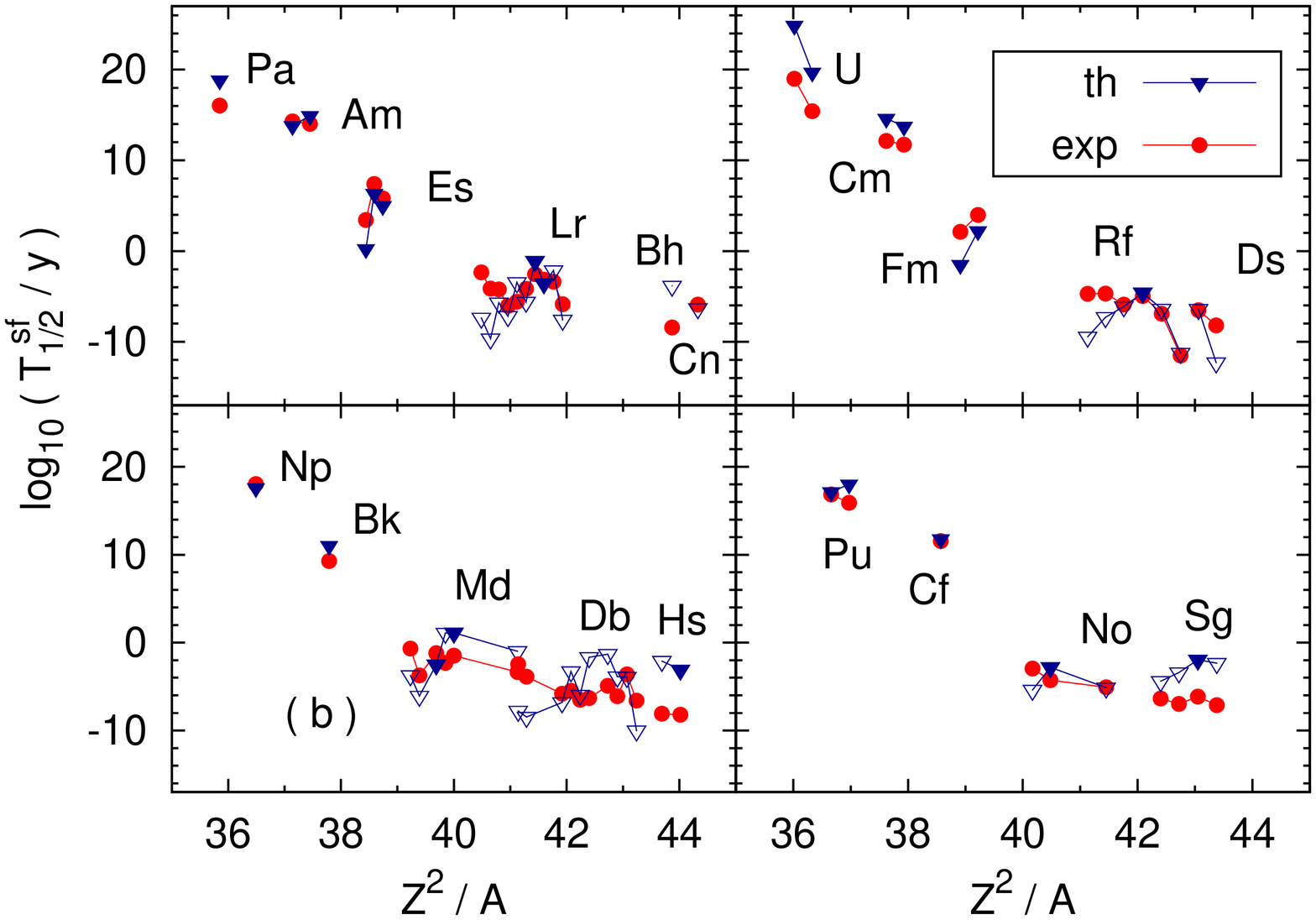}
\caption{Spontaneous fission half-lives of even-even (top) and odd (bottom)
nuclei, calculated using formula (\ref{Tsf}) in comparison to the experimental
values \cite{nudat2}. The open symbols correspond to the estimates evaluated on
the basis of extrapolated experimental masses in Ref. \cite{nudat2}.}
\label{fig3}
\end{figure}

Using the approximation (\ref{fZ}) one can estimate the spontaneous fission
half-lives by the following formula:
\begin{equation}\label{Tsf}
\begin{array}{ll}
\log_{10}\left(\frac{T_{1/2}^{\rm sf}(Z,N)}{y}\right)&
    =-4 Z+371.2-7.5\delta M(Z,N)\\
&+\left\{
\begin{array}{cl}
0   &{\rm ~for~even-even \,\,,}\\
10.3   &{\rm ~for~odd-}A\,\,,\\
17.8&~ {\rm for~odd-odd\,\,,}
\end{array}\right.
\end{array}
\end{equation}
where the experimental microscopic correction $\delta M$ is defined in
Eq.~(\ref{dM}). The estimates of $T_{1/2}^{\rm sf}$ obtained with formula
(\ref{Tsf}) are compared in Fig.~\ref{fig3} with the experimental data taken
from Ref.~\cite{nudat2}.

Surprisingly good agreement of the model estimates with the data is achieved for
all actinide and even super-heavy nuclei. The root mean square deviation of
$\log T_{1/2}^{\rm sf}$ for the even-even isotopes with $Z< 104$ is 1.52 and it
grows to 2.02 when odd-$A$ and odd-odd nuclei are taken into account. The rms
deviation reaches 2.47 when one includes to the analysis the super-heavy
isotopes (for which the liquid drop barrier vanishes). To foresee $T_{1/2}^{\rm
sf}$ for unknown nuclei one can evaluated $\delta M$ using mass estimates from
one of existing on the market mass tables.

In order to understand this striking for the first sight result it is good to
remind the topographical theorem, proposed by Myers and \'Swi{}\c{a}tecki
\cite{MS96}: the mass of a nucleus in a saddle point is determined by the
macroscopic part of the binding energy. The shell effects at the saddle are
negligible, so the fission barrier heights is approximately determined by the
difference of the macroscopic (here LSD) and the experimental masses:
\begin{equation}
 V_{\rm B}(Z,N) = M_{\rm LSD}(Z,N,{\rm saddle}) - M_{\rm exp}^{g.s.}.
\end{equation}
The function $F(Z)$ (\ref{fZ}) represents the liquid-drop estimate of the
fission life-time and the $k$ parameter in Eq.~(\ref{Tsf}) gives enhancement
of the life-time due to increase of the fission barrier due to the ground state
microscopic effects. This parameter $k$ plays a similar role as the hindrance
factor $h$ which origins from the barrier augmentation due to spin and parity
conservation of an odd-$A$ nucleus along the path to fission.

\section{Summary}

The following conclusions can be drawn from our investigation:\\[-5ex]
\begin{itemize}
\item Simple phenomenological formula for the spontaneous fission half-lives
depending on proton number and microscopic energy correction in the ground
state was found.\\[-5ex]
\item Model estimates of the  spontaneous fission half-lives of even-even nuclei
are in a surprisingly good agreement with experimental data.\\[-5ex]
\item Quality of evaluation for odd nuclei is much worse, what is due to fact
that the effect of an odd-particle on the barrier penetrability is described
here by a single constant, independent on angular momentum or parity of the odd
nucleon.\\[-5ex]
\item Our simple formula with the microscopic energy corrections taken e.g. from
the tables \cite{MN95} can serve to rough estimates of the fission life-times of
new undiscovered yet isotopes.
\end{itemize}
The present phenomenological formula for the spontaneous fission half-lives
and similar formula for the alpha decay half-lives derived in Ref.~\cite{ZW13}
can serve a useful tool to estimate stability of non discovered yet isotopes.
The only input which one needs in the both formulas are the binding energy
estimates which can be taken from one of the mass model existing on the
market.

\section*{Acknowledgements}

 This work was partly supported by the Polish National Science Center Grant
 No.~DEC-2011/01/B/ST2/03667.

\end{document}